\documentclass[aip, jcp, preprint]{revtex4-1}
\usepackage[utf8]{inputenc}
\usepackage[english]{babel}
\usepackage{color}
\usepackage[margin = 1.0in]{geometry}
\usepackage{amsmath}
\usepackage{amssymb}
\usepackage{graphicx}
\usepackage{url}
\usepackage[colorlinks=true, allcolors=blue]{hyperref}
\usepackage{mathtools, nccmath}
\usepackage{float}
\usepackage{mathrsfs}
\usepackage{yfonts}
\usepackage{array}
\usepackage{braket}

\usepackage{feynmp}
\usepackage{caption}
\usepackage{subcaption}
\usepackage{natbib}
\usepackage{tikz}
\usepackage{bm}
\DeclareUnicodeCharacter{2212}{\hspace{1cm}-}

\begin{document}

\title{Stabilised Coupled Trajectory Mixed Quantum Classical Algorithm with Improved Energy Conservation: CTMQC-EDI}

\author{Aaron Dines\textsuperscript{1}, Matthew Ellis\textsuperscript{1} and Jochen Blumberger\textsuperscript{}}
\affiliation{ \textsuperscript{\textnormal{1}} Department of Physics and Astronomy and Thomas Young Centre, University College London, Gower Street, London WC1E 6BT, United Kingdom}

\date{October 2023}

\begin{abstract}
Coupled trajectory mixed quantum classical (CTMQC) dynamics is a rigorous approach to trajectory-based non-adiabatic dynamics, 
which has recently seen an improvement to energy conservation via the introduction of the CTMQC-E algorithm. Despite this, the method's 
two key quantities distinguishing it from Ehrenfest dynamics, the modified Born-Oppenheimer momentum and the 
quantum momentum, require regularisation procedures in certain circumstances. Such procedures in the latter can cause instabilities leading to undesirable effects such as energy drift and spurious population transfer, which is expected to become increasingly prevalent the larger the system as such events would happen more frequently. We propose a further modification to CTMQC-E which includes a redefinition of the quantum momentum, 
CTMQC-EDI (Double Intercept), such that it has no formal divergences. We then show for Tully models I-IV that the algorithm has greatly improved total energy conservation and 
negligible spurious population transfer at all times, in particular in regions of strong non-adiabatic coupling. CTMQC-EDI therefore shows promise as 
a numerically robust non-adiabatic dynamics technique that accounts for decoherence from first principles and that is scalable to large molecular systems and materials.  
\end{abstract}

\maketitle



\section{Introduction}
The treatment of electronically non-adiabatic phenomena such as 
photoexcitation\cite{doi:10.1021/acs.chemrev.9b00447} and charge transport in molecules and materials\cite{doi:10.1021/cr050140x, doi:10.1021/acs.accounts.1c00675}
require approaches that go beyond the Born-Oppenheimer approximation\cite{born2000quantum}. 
Fully quantum dynamical non-adiabatic methods are available at various levels of approximation, ranging from Multi-Configurational Time-Dependent Hartree (MCTDH)\cite{MCTDH_orig} to 
ab initio multiple spawning (AIMS)\cite{AIMS_orig} and ab initio multiple cloning (AIMC)\cite{AIMC_orig}, but they become computationally prohibitive for systems with a large 
number of degrees of freedom, e.g., materials. In this case one often resorts to quantum-classical non-adiabatic dynamics methods where electrons are treated quantum mechanically 
and nuclei as classical particles with some of their quantum character incorporated by simulating a swarm of classical trajectories. 


Various quantum-classical procedures have been proposed and successfully implemented, most prominently Ehrenfest dynamics\cite{ehrenfest1927bemerkung} and Tully Trajectory Surface Hopping (SH)\cite{tully_model}. However, both methods have well-documented deficiencies. Most problematic is the overcoherence of the electronic wavefunction causing the electronic temperature to tend to infinity at long times. Decoherence-corrected Ehrenfest and surface hopping schemes remedy this problem often very successfully, but these are ad hoc correction schemes not derived from first principles and although they often show excellent performance for trajectory-ensemble averages, the details of the underlying electron-nuclear dynamics may not be very accurate. Recently, a spin mapping approach to surface hopping (MASH) has been proposed which has deterministic trajectories and a natural decoherence mechanism\cite{MASH}. In this theory, decoherence is realised through resampling techniques known as quantum-jumps; they are rigorously applicable because of MASH's connection to the short-time Quantum Classical Liouville Equation (QCLE)\cite{QCLE}.  Although initially restricted to two state systems, it has been successfully generalised to multi-state systems\cite{multi-MASHF2016}. Results from this method appear promising, as they can converge to full quantum mechanics results in some cases.




The correct description of electron-nuclear dynamics, especially as it effects decoherence, can be central to our interpretations of what is happening in non-adiabatic systems. It is therefore desirable to have an algorithm which can incorporate these effects which is understood from first principles, and retains its computational advantage over full quantum dynamical methods. Coupled Trajectory Mixed Quantum Classical (CTMQC) dynamics \cite{orig_ctmqc} may be viewed as an extension of Ehrenfest dynamics\cite{ehrenfest1927bemerkung}, which includes a first-principles approach to the decoherence problem as it is derived from the exact factorisation\cite{exact_factorisation} of the wavefunction into nuclear and electronic subsystems. The algorithm is obtained following a number of well-defined approximations, principally by taking the classical limit of the equation of motion for the nuclear wavefunction. Trajectories become coupled as a consequence of the quantum momentum,  which depends on the gradient of the square-root of the nuclear density, which is then approximated from the swarm of classical trajectories. CTMQC is thus, in principle, a systematically improvable method, in contrast to SH. 

CTMQC has been tested extensively on model systems\cite{gossel_benchmark, doi:10.1021/acs.jctc.0c00679, energy_expression, ct_tsh} and even on higher dimensional systems\cite{doi:10.1021/acs.jpclett.7b01249}. In these implementations, CTMQC excellently captures the populations of and coherence between electronic states, often outperforming methods such as SH-EDC (Energy Decoherence Correction)\cite{SH-EDC} if the errors from each metric are considered simultaneously\cite{ct_tsh}. Additionally, the fact that the theory is derivable provides a bridge between itself and full quantum mechanics allowing for systematic improvements to the algorithm; a CTMQC algorithm capable of including quantum nuclear tunneling has been proposed and implemented for a spin-boson model and the improvement was shown to have excellent agreement with exact quantum mechanics, especially pertaining to the transmission probabilities of trajectories and the evolution of the nuclear wavepacket. Notably, this improvement was attained by simply lifting the assumption that the quantum potential is negligible, and by approximating it using the nuclear wavepacket reconstructed from coherent states centred on the classical trajectories\cite{CTMQC-NQE}.



The original CTMQC algorithm suffers from two key limitations: energy non-conservation and numerical instabilities arising from the divergence of the quantum momentum. The former of these problems has been addressed with the modified algorithm CTMQC-E, which was shown to improve energy conservation substantially in Tully model III (Extended Coupling Region model) and in a retinal chromophore model\cite{ctmqc-e}. Moreover, it has recently been shown that the improvements of CTMQC-E compared with CTMQC for more complex model systems than the Tully models has a large impact on the accuracy of the methods in reproducing electronic populations and nuclear densities \cite{unpublished_rev}. 
However, energy conservation is only conserved in regions where the kinetic energy of a trajectory is non-zero, and where an expression for the quantum momentum that prevents spurious population transfer can be used (see below for a definition and detailed discussion of spurious population transfer). If either one of these conditions is not met then energy drift will occur. Although this may not have a significant impact on low-dimensional systems, large systems with thousands of degrees of freedom run over picosecond timescales, such as would be typically be simulated for nanoscale systems, will accumulate large energy deviations and thus unphysical dynamics is likely to happen.  The dynamics are also impacted by the treatment of the quantum momentum even when energy conservation is well-satisfied. We aim to address these problems by investigating how the quantum momentum's numerical instabilities are treated.

Here, we propose a redefinition of the quantum momentum which we refer to as the Double Intercept (DI) method, which removes divergences formally and greatly reduces spurious population transfer at all times, especially
in regions of strong non-adiabatic coupling (NAC), where current implementations fail to do so. 
We show that this results in a marked improvement to the energy conservation in all of the Tully Models (I-IV) for low and high momentum cases, and explain why this is a generic feature of the algorithm that 
we expect to extend to more complex systems. This method is then compared with the Cut-Off procedure similar to the one used in the g-CTMQC code \cite{g_CTMQC}, and another novel regularisation procedure which we introduce in this work.

This paper is structured as follows: in section II we first provide the theory of CTMQC with how one arrives at the quantum momentum expression requiring regularisation; 
we then provide a brief overview of CTMQC-E, and then we introduce the methods to address the divergences. After Computational Details in section III, we present the  
results of our simulations in Section IV, in particular for  DI with E: CTMQC-EDI. Finally, in section V, a summary is given to appraise the methods and suggest further work.

\section{Theory}
\subsection{CTMQC}
CTMQC is a deterministic coupled trajectory-based mixed quantum classical scheme wherein electronic dynamics follow extended 
Ehrenfest equations, with novel terms in the equations of motion arising from the coherence of electronic states and the quantum 
momentum. Specifically, the evolution of the coefficients of the electronic wavefunction in the Born-Huang expansion, 
$\Phi_{\bf{R}} ({\bf r}, t)  = \sum_m C_m (\mathbf{R},t) \phi_{\mathbf{R},m} ({\bf r})$ (where $\mathbf{R}$, $\mathbf{r}$ denote nuclear and electronic positions and 
$\phi_{\mathbf{R},m}$ the Born-Oppenheimer or adiabatic electronic state eigenfunctions)\cite{ct_tsh}, may be split into Ehrenfest (Eh) and exact factorisation (XF) contributions,
\begin{equation}\label{electronic_equation_split}
    \frac{d}{dt}C_{m}^{(\alpha)}(t) = \frac{d}{dt}C_{m, Eh}^{(\alpha)}(t)+\frac{d}{dt}C_{m, XF}^{(\alpha)}(t),
\end{equation}
defined by \cite{ctmqc-e},
\begin{equation}\label{Ehrenfest_contribution}
\begin{split}
    \frac{d}{dt}C_{m, Eh}^{(\alpha)}(t) \equiv \frac{-i}{\hbar} E_{m}^{(\alpha)}(t)C_{m}^{(\alpha)}(t) - \sum_{l}\sum_{\nu=1}^{N_{n}}\dot{\mathbf{R}}_{\nu}^{(\alpha)}(t)\cdot\mathbf{d}_{\nu,ml}^{(\alpha)}C_{l}^{(\alpha)}(t),   
\end{split}
\end{equation}
and,
\begin{equation}\label{XF_contribution}
\begin{split}
\frac{d}{dt}C_{m, XF}^{(\alpha)}(t)\equiv\sum_{l}\sum_{\nu=1}^{N_{n}}\frac{\bm{\mathcal{P}}_{\nu}^{(\alpha)}(t)}{\hbar M_{\nu}}\cdot (\mathbf{f}_{\nu,m}^{(\alpha)}(t) - \mathbf{f}_{\nu,l}^{(\alpha)}(t))C_{l}^{(\alpha)}(t)\rho_{lm}^{(\alpha)}(t). 
\end{split}    
\end{equation}

Where $\mathbf{R}^{(\alpha)}_{\nu}(t)$ and $\dot{\mathbf{R}}^{(\alpha)}_{\nu}(t)$ are the position and velocity of nucleus $\nu$ with mass $M_\nu$
on classical trajectory $\alpha$, $E_{m}^{(\alpha)}$ is the Born-Oppenheimer (or adiabatic) potential energy for electronic state $m$ on trajectory $\alpha$, 
$\mathbf{d}_{\nu,ml}^{(\alpha)}$ is the non-adiabatic coupling vector between the adiabatic electronic states $m$ and $l$, 
$\rho_{lm}^{(\alpha)}(t)\equiv C_{l}^{(\alpha) *}(t) C_{m}^{(\alpha)}(t)$ is the electronic density matrix and 
$\bm{\mathcal{P}}_{\nu}^{(\alpha)}(t) = -\hbar\nabla_{\nu}\ln{|\chi(\mathbf{R},t)|}$  is the quantum momentum 
corresponding to the nuclear wavefunction $\chi(\mathbf{R},t)$. The XF contributions in Eq.~\eqref{XF_contribution} also contain what is 
referred to as the Born-Oppenheimer momentum, $\mathbf{f}^{(\alpha)}_{\nu,m}(t)$, since in the original algorithm it is approximated 
via $\mathbf{f}_{\nu,m}^{(\alpha)}(t) \approx \int^{t}-\nabla_{\nu}E_{m}^{(\alpha)}d\tau$. Note that we distinguish when 
a quantity is only dependent on the nuclear configurations, as with the BO energies, and when it is also dependent explicitly 
on time, as with the BO momentum. 

The Ehrenfest contributions in Eq.~\eqref{Ehrenfest_contribution} lead to mixing of, and population transfer between, adiabatic states 
in regions of strong NAC (second term on the right-hand side (RHS) of Eq.~\eqref{Ehrenfest_contribution}) whilst the XF contribution in 
Eq.~\eqref{XF_contribution} is responsible for the decoherence of individual trajectories to adiabatic states 
once the nuclei have left regions of strong NAC. Here we define coherence as the modulus square of the off-diagonal elements of the electronic
density matrix, $|\rho_{lm}^{(\alpha)}(t)|^2  = |C_{l}^{(\alpha)}(t)|^2 |C_{m}^{(\alpha)}(t)|^2$, $l \neq m$. 
Since the rate of change of the populations in such regions 
will be dependent on the coherence between the states, it is intuitive that 
equilibrium populations are only reached when coherence terms vanish. 

Nuclear forces are derived from the classical limit of the nuclear evolution equation appearing in the exact factorisation of the time-dependent Schr\"{o}dinger equation\cite{orig_ctmqc, exact_factorisation}. Importantly, the forces derive directly from the electronic wavefuntion via a Berry vector potential. Once again these contributions to the force can be split into (Eh) and (XF) contributions\cite{ctmqc-e},

\begin{equation}\label{force_decomp}
    \mathbf{F}_{\nu}^{(\alpha)}(t) = \mathbf{F}_{\nu,Eh}^{(\alpha)}(t)+\mathbf{F}_{\nu,XF}^{(\alpha)}(t),
\end{equation}
with the respective definitions,
\begin{equation}\label{Ehrenfest_force}
\begin{split}
    \mathbf{F}_{\nu,Eh}^{(\alpha)}(t) &\equiv -\sum_{m}\rho_{mm}^{(\alpha)}(t)\nabla_{\nu}E_{m}^{(\alpha)} + \sum_{m,l}\rho_{ml}^{(\alpha)}(t)(E_{m}^{(\alpha)}-E_{l}^{(\alpha)})\mathbf{d}^{(\alpha)}_{\nu,ml}C_{l}^{(\alpha)}(t),
\end{split}
\end{equation}
and,
\begin{equation}\label{XF_force}
\begin{split}
    \mathbf{F}_{\nu,XF}^{(\alpha)}(t)&\equiv\sum_{m,l}\left[\sum_{\mu = 1}^{N_{n}}\frac{\bm{\mathcal{P}}_{\mu}^{(\alpha)}(t)}{\hbar M_{\mu}}\cdot\left(\mathbf{f}_{\mu,m}^{(\alpha)}(t)-\mathbf{f}_{\mu,l}^{(\alpha)}(t)\right)\right]\times|\rho_{ml}^{(\alpha)}(t)|^{2}(\mathbf{f}^{(\alpha)}_{\nu,m}(t)-\mathbf{f}^{(\alpha)}_{\nu,l}(t)).
\end{split}
\end{equation}
All XF based terms are proportional to the electronic coherence matrix and thus only contribute when the coherence between states on a given trajectory is non-zero. 


From Eqs.~\eqref{XF_contribution}-\eqref{XF_force} it is clear that an accurate approximation for the quantum momentum is central to the success of the algorithm. 
However, during simulations we do not have access to the nuclear density directly. One approximation is to reconstruct the nuclear density from the swarm of classical 
trajectories using Gaussians centred on the classical nuclei. The corresponding 
quantum momentum is a quasi-linear function of the nuclear position containing a slope $\frac{\hbar}{2\sigma_{\nu}(t)^{2}}$ and an intercept term ${\bm{\mathcal{Y}}}_{\nu}^{(\alpha)}(t)$, 
\begin{equation}\label{intercept}
{\bm{\mathcal{P}}}^{(\alpha)}_{\nu}(t) = \frac{\hbar}{2\sigma_{\nu}(t)^{2}} {\bf{R}}_{\nu}^{(\alpha)}(t) - {\bm{\mathcal{Y}}}_{\nu}^{(\alpha)}(t),
 \end{equation}
where for an explicit expression of the intercept in terms of the Gaussians centred on the nuclei we refer to Ref.  \cite{doi:10.1021/acs.jpclett.7b01249}. $\sigma_{\nu}(t)$ is the width of the Gaussian centred on nucleus $\nu$. It may be treated as time-independent 
in the so-called ``frozen Gaussian" approximation\cite{FG} or as time and trajectory-dependent ($\sigma^{(\alpha)}_{\nu}(t)$), but we neglect 
this detail to simplify our arguments; the same conclusions apply in the latter, more general case. We do however note that the width may vary in each dimension independently, so we later use $\sigma_{j}$ to refer to a width in cartesian direction $j\in3N_{n}$. 

The expression for the intercept term derived as described above is known to cause population transfer on average between adiabatic states even in regions where the NACs are zero. This is an undesirable result that is not reflected in full quantum dynamics simulations. To remedy this problem the following condition is imposed in 
CTMQC \cite{gossel_benchmark, ct_tsh, ctmqc-e},
\begin{equation}\label{zero_pop_transfer}
\begin{split}
    {\text{if}}\,\, \mathbf{d}_{\nu,ml}^{(\alpha)}=0 \hspace{0.1cm}\forall\,\,\nu,m,l,\alpha \rightarrow \Big\langle\frac{d}{dt}|C^{(\alpha)}_{m}(t)|^{2}\Big\rangle_{(\alpha)} = 0 \hspace{0.1cm}\forall\,\,m, 
\end{split}
\end{equation}
where $\langle...\rangle_{(\alpha)}$ refers to an average over all trajectories. We refer to this as the spurious population transfer condition. 
The condition may be considered an intuitive consequence of the restoration of the adiabatic approximation in regions where the NACs are close to zero.
Inserting the condition, Eq.~\eqref{zero_pop_transfer} in Eqs.~\eqref{electronic_equation_split}-\eqref{XF_contribution}, one obtains a more explicit form of the 
spurious transfer condition, 
\begin{equation}\label{zero_pop_transfer_2}
\begin{split}
&\left\langle \sum_{l}\sum_{\nu=1}^{N_{n}}\frac{2\bm{\mathcal{P}}_{\nu}^{(\alpha)}(t)}{\hbar M_{\nu}}\cdot (\mathbf{f}_{\nu,m}^{(\alpha)}(t) - \mathbf{f}_{\nu,l}^{(\alpha)}(t)) 
|\rho_{lm}^{(\alpha)}(t)|^2\right\rangle_{(\alpha)} = 0  \hspace{0.1cm}\forall\,\,m,  
\end{split}
\end{equation}
where the quantum momentum $\bm{\mathcal{P}}_{\nu}^{(\alpha)}$ is given by Eq.~\eqref{intercept}. 
%
%
After asserting that the condition of zero population transfer, Eq.~\eqref{zero_pop_transfer_2}, is separately zero for each pairwise 
contribution from adiabatic states $(m,l)$ as well as for each degree of freedom $j \in 3N_{n}$, accompanied by a pairwise and 
trajectory independent intercept for each degree of freedom, ${\bm{\mathcal{Y}}}_{\nu}^{(\alpha)}\rightarrow \mathcal{Y}_{j,ml}$,  
the spurious transfer condition becomes
\begin{equation}\label{original_pop_transfer}
    \Big\langle\mathcal{P}^{(\alpha)}_{j,ml}(t)\left(f^{(\alpha)}_{j,m}(t)-f^{(\alpha)}_{j,l}(t)\right)|\rho^{(\alpha)}_{ml}(t)|^{2}\Big\rangle_{(\alpha)} = 0\hspace{0.1cm}\forall\,\, m,l,j  
\end{equation}
and the expression for the quantum momentum is obtained as\cite{doi:10.1021/acs.jpclett.7b01249, ct_tsh}:
\begin{equation}\label{traditional_quantum_momentum}
    \mathbf{\mathcal{P}}^{(\alpha)}_{j,ml}(t) = \frac{\hbar}{2\sigma_{j}(t)^{2}}R^{(\alpha)}_{j}(t) - \mathcal{Y}_{j,ml}(t),
\end{equation}
with the intercept term uniquely determined as,
\begin{equation}\label{old_y-intercept}
  \begin{split} 
    &\mathcal{Y}_{j,ml}(t) = \frac{\hbar}{2\sigma_{j}(t)^{2}}\frac{\Big\langle R_{j}^{(\alpha)}(t)\left(f^{(\alpha)}_{j,m}(t) -f^{(\alpha)}_{j,l}(t)\right)|\rho^{(\alpha)}_{ml}(t)|^{2}\Big\rangle_{(\alpha)}}{\Big\langle\left(f^{(\beta)}_{j,m}(t) -f^{(\beta)}_{j,l}(t)\right)|\rho^{(\beta)}_{ml}(t)|^{2}\Big\rangle_{(\beta)}}.
  \end{split}  
\end{equation}



The above parameterisation breaks the definition of the quantum momentum as a logarithmic derivative of the nuclear density. 
The main problem is however in the nature of the intercept term Eq.~\eqref{old_y-intercept}, since it is divergent if the denominator 
tends to zero and the numerator remains finite.   
%
This is where the instabilities in the algorithm can arise. Unfortunately, this is not an uncommon occurrence, as in our analysis of the Tully 
models we found that all models encountered these instabilities, apart from a few cases, such as the high momentum case in Tully model III (Extended Coupling Region)\cite{tully_model}.

To address these problems a Cut-Off procedure is commonly used. For example the g-CTMQC code\cite{g_CTMQC} employs the following criteria: the intercept term is compared to the position of a trajectory and if this trajectory is more than a certain number of standard deviations, $\sigma_{j}(t)$, from the intercept then the following definition is used instead\cite{doi:10.1021/acs.jpclett.7b01249},
\begin{equation}\label{gaussian_intercept}
    \mathcal{Y}_{j}^{(\alpha)}(t) = \frac{\hbar}{2\sigma_{j}(t)^{2}}\frac{\sum_{\beta=1}^{N_{traj}}R^{(\beta)}_{j}(t)G_{\sigma}(\mathbf{R}^{(\alpha)}(t)-\mathbf{R}^{(\beta)}(t))}{\sum_{\gamma=1}^{N_{traj}}G_{\sigma}(\mathbf{R}^{(\alpha)}(t)-\mathbf{R}^{(\gamma)}(t))},
\end{equation}
where the $G_{\sigma}(\mathbf{R}^{(\alpha)}(t)-\mathbf{R}^{(\beta)}(t))$ represents the multidimensional Gaussian with standard deviation $\sigma$ centred at $\mathbf{R}^{(\beta)}(t)$. If this definition is also more than the chosen number of standard deviations away then the quantum momentum is set to zero. The procedure is based on the fact that a trajectory far enough away from all the others can be considered as independent and thus would not experience decoherence effects due to the branching of trajectories unless it came into contact with them again later in the simulation. The procedure also defines an absolute cut-off to the value of the quantum momentum which is given by plugging this threshold criterion into the expression for the quantum momentum in Eq. \eqref{traditional_quantum_momentum}, which gives $\mathcal{P}_{\textrm{cut-off}, j}(t) = \frac{10\hbar}{2\sigma_{j}(t)}$. The key drawback is that 
the quantum momentum with intercept the intercept in Eq.~\eqref{gaussian_intercept} no longer in general satisfies the spurious population transfer condition of Eq.~\eqref{zero_pop_transfer}. This gives rise to fairly large energy drifts, as we will show below, even if the total energy correction scheme (CTMQC-E) is applied.


\subsection{CTMQC-E}

The quantum-classical expression for the energy on a single trajectory is given by the sum of the population weighted adiabatic energies and the kinetic energy of the classical nuclei\cite{energy_expression,ctmqc-e}:
\begin{equation}\label{energy_expression}
    E^{\alpha} = E^{(\alpha)}_{kin} + \sum_{m}\rho_{mm}^{(\alpha)}(t)E_{m}^{(\alpha)},
\end{equation}
where $E^{(\alpha)}_{kin} \equiv \sum_{\nu=1}^{N_n}\frac{1}{2}M_{\nu}|\dot{\mathbf{R}}_{\nu}^{(\alpha)}|^{2}$. The conserved quantity is the trajectory average,
\begin{equation} \label{energy_conservation}
\frac{d}{dt} \langle E^{(\alpha)} \rangle_{(\alpha)}\!=\!0. 
\end{equation}
This is more faithful to the Ehrenfest theorem in full quantum mechanics, by contrast to applying energy conservation on each trajectory independently\cite{ctmqc-e}. 
Eq.~\eqref{energy_conservation} leads to a sum of expressions, some of which derive from Ehrenfest, 
and some of which arise purely from the XF-based terms. It can be shown that since the Ehrenfest contribution 
to the force derives from the gradient of the potential energy expression, then the total time derivative of the Ehrenfest 
contributions will cancel each other analytically on each trajectory. Therefore the analytical contribution to energy 
deviation is the contribution from the XF terms solely:
\begin{equation}\label{Energy_deviation_XF}
\begin{split}
    &\hspace{-0.6cm}\langle\dot{E}\rangle =  
    \Bigg\langle\sum_{m,l}\sum_{j=1}^{3N_{n}}\frac{\mathcal{P}_{j,ml}^{(\alpha)}(t)}{\hbar M_{j}}|\rho_{ml}^{(\alpha)}(t)|^{2}\left(f_{j,m}^{(\alpha)}(t)-f_{j,l}^{(\alpha)}(t)\right)
    \\
    &\hspace{-0.6cm}\times\left[\sum_{k=1}^{3N_{n}}\left(f_{k,m}^{(\alpha)}(t)-f_{k,l}^{(\alpha)}(t)\right)\dot{R}^{(\alpha)}_{k}(t) + \left(E^{(\alpha)}_{m}-E^{(\alpha)}_{l}\right)\right]\Bigg\rangle_{(\alpha)}. 
\end{split}
\end{equation}
The strategy employed by the CTMQC-E algorithm is to make the above expression formally equivalent
to Eq.~\eqref{zero_pop_transfer}, and when the latter is expanded into its component form, as above, we see that if the product on the second line of Eq.~\eqref{Energy_deviation_XF}
is trajectory independent then the expressions will be equivalent up to a time- dependent function. Therefore Eq.~\eqref{old_y-intercept} will prevent 
both spurious population transfer and energy drift incurred by the quantum momentum terms. The value which is chosen for the product on the second line of Eq.~\eqref{Energy_deviation_XF} 
is arbitrary but in each case the value is achieved by redefining the BO momenta $f^{(\alpha)}_{j,m}(t)\rightarrow \tilde{f}^{(\alpha)}_{j,m}(t)$. If one chooses it to be equal to its trajectory average with respect to the original BO momenta and simultaneously chooses the redefined BO momentum to be parallel to the velocity, then this will yield the
BO momentum presented in the CTMQC-E paper\cite{ctmqc-e}:
\begin{equation}\label{redef_BO_momentum}
\begin{split}
    &\tilde{\mathbf{f}}^{(\alpha)}_{\nu,m}(t) = \frac{\left(-E_{m}^{(\alpha)}+\left\langle\left[\sum_{k=1}^{3N_{n}}f^{(\beta)}_{k,m}(t)\dot{R}^{(\beta)}_{k}(t) + E^{(\beta)}_{m}(t)\right]\right\rangle_{(\beta)}\right)}{2E_{kin}^{(\alpha)}}\times M_{\nu}\dot{\mathbf{R}}_{\nu}^{(\alpha)}(t).
\end{split}
\end{equation}
Clearly this redefinition is ill-defined on trajectories with zero kinetic energy and thus a cut-off is used for the denominator, below which the original definition of the BO momentum is used\cite{ctmqc-e}. 

\subsection{CTMQC-DI and CTMQC-EDI}
The way in which divergences of the intercept term of Eq.\eqref{old_y-intercept} are handled in the Cut-Off method is known to cause instabilities in CTMQC and CTMQC-E. 
Here we define two novel procedures to address this: the Double Intercept (DI) method and Regularisation (R). 

In the DI method we consider the spurious population transfer condition Eq.~\eqref{original_pop_transfer} 
but we now assert a stricter condition wherein the population transfer is forbidden on average on the subsets of trajectories where 
$f^{(\alpha)}_{j,m}(t)-f^{(\alpha)}_{j,l}(t)<0$ and $f^{(\alpha)}_{j,m}(t)-f^{(\alpha)}_{j,l}(t)>0$ separately,
\begin{equation}\label{zero_pop_transfer_plus}
    \Big\langle\mathcal{P}^{(\alpha)}_{j,ml}(t)\left(f^{(\alpha)}_{j,m}(t)-f^{(\alpha)}_{j,l}(t)\right)|\rho^{(\alpha)}_{ml}(t)|^{2}\Big\rangle^{(+)}_{(\alpha)} = 0\hspace{0.1cm}\forall\,\, m,l,j
\end{equation}
and
\begin{equation}\label{zero_pop_transfer_minus}
    \Big\langle\mathcal{P}^{(\alpha)}_{j,ml}(t)\left(f^{(\alpha)}_{j,m}(t)-f^{(\alpha)}_{j,l}(t)\right)|\rho^{(\alpha)}_{ml}(t)|^{2}\Big\rangle^{(-)}_{(\alpha)} = 0\hspace{0.1cm}\forall\,\, m,l,j
\end{equation}
where the superscript $(\pm)$ denotes the subset. The new intercepts are given respectively as,
\begin{equation}\label{DI}
    \mathcal{Y}^{(\pm)}_{j,ml}(t) = \frac{\hbar}{2\sigma_{j}(t)^{2}}\frac{\Big\langle R^{(\alpha)}_{j}(t)\left(f^{(\alpha)}_{j,m}(t) -f^{(\alpha)}_{j,l}(t)\right)|\rho^{(\alpha)}_{ml}(t)|^{2}\Big\rangle^{(\pm)}_{(\alpha)}}{\Big\langle\left(f^{(\beta)}_{j,m}(t) -f^{(\beta)}_{j,l}(t)\right)|\rho^{(\beta)}_{ml}(t)|^{2}\Big\rangle^{(\pm)}_{(\beta)}}.
\end{equation}
The above equation is the key analytical result of the paper.  It may be coupled to CTMQC-E by replacing the BO momenta with their modified expressions: $f_{j,m}^{(\alpha)}(t)\rightarrow\tilde{f}_{j,ml}^{(\alpha)}(t)$, defined in
Eq.~\eqref{redef_BO_momentum}. We refer to the resulting algorithm as CTMQC-EDI.

In the DI method divergences are resolved because on each subset the weight factor $w_{j,ml}^{(\alpha), (\pm)}(t)\equiv\frac{1}{N_{traj}}\frac{\left(f^{(\alpha)}_{j,m}(t) -f^{(\alpha)}_{j,l}(t)\right)|\rho^{(\alpha)}_{ml}(t)|^{2}}{\Big\langle\left(f^{(\beta)}_{j,m}(t) -f^{(\beta)}_{j,l}(t)\right)|\rho^{(\beta)}_{ml}(t)|^{2}\Big\rangle^{(\pm)}_{(\beta)}}$ is finite for all values of its denominator. In fact the weight factor takes values $w_{j,ml}^{\alpha,(\pm)}(t)\in [0,1], \forall\,\, \alpha$  with the property that the weights sum to unity. Thus we have also defined a natural cut-off to each branch of the quantum momentum which is given by $\max\left(|\mathcal{P}^{{(\alpha)},(\pm)}_{j,ml}(t)|\right)=\max\left(\frac{\hbar}{2\sigma_{j}(t)^{2}}|R_{j}^{(\alpha)}(t)-R_{j}^{(\pm)}(t)|\right)$, where $R^{(\pm)}_{j}(t)$ are the set of nuclear positions on the respective subsets. Therefore the sum which defines the intercept term on each is always finite.

The prescription of $(\pm)$ subsets depends on the ordering of the indices $m,l$. When one reverses the ordering $m\leftrightarrow l$ one should also take the superscript $(\pm)\rightarrow(\mp)$. Thus the intercept term has the property $Y_{j,ml}^{(\pm)}=Y_{j,lm}^{(\mp)}$. A given trajectory, $\alpha$, and direction, $j$, can belong to the positive or negative branch depending on the index ordering, but this has no impact on the numerical value of the quantum momentum; we still have the property that $\mathcal{P}^{(\alpha)}_{j,ml}(t)=\mathcal{P}_{j,lm}^{(\alpha)}(t)\hspace{0.1cm} \forall\,\,\alpha,j,l,m$. 

If a trajectory has a BO momentum difference of zero, then it is not included in either calculation of the intercepts. These trajectories in fact follow Ehrenfest dynamics exactly, as can be seen from Eq.~\eqref{XF_contribution} for the evolution of adiabatic coefficients and Eq.~\eqref{XF_force} for the nuclear forces. Any choice of quantum momentum in these regions will yield the same dynamics so we set the quantum momentum to zero here. 


The quantum momentum can discontinuously change upon switching branches, however this nonetheless analytically results in continuous forces and evolution of adiabatic states as the force expression \eqref{XF_force} and electronic evolution \eqref{XF_contribution} rely only on $\mathcal{P}^{(\alpha)}_{j,ml}(f^{(\alpha)}_{j,m}(t)-f^{(\alpha)}_{j,l}(t))$ being continuous in time. Which is generally satisfied because in order for a trajectory to switch branch it must go through $(f^{(\alpha)}_{j,m}(t)-f^{(\alpha)}_{j,l}(t)) = 0$ continuously. Therefore the new prescription for the quantum momentum in CTMQC-DI and CTMQC-EDI is well defined and no longer causes divergences in wavefunction evolution or 
force calculation.         


\subsection{CTMQC-R and CTMQC-ER}
An alternative way to remove divergences in the intercept term in Eq.\eqref{old_y-intercept} is to apply a regularisation (R) procedure. We define the dimensionless function,
\begin{equation}\label{regularisation_1}
\begin{split}
    \hspace{-0.4cm}x_{j,ml}^{(\alpha)}(t) \equiv\frac{\sigma_{j}(t)\Big\langle\left(f^{(\gamma)}_{j,m}(t) -f^{(\gamma)}_{j,l}(t)\right)|\rho^{(\gamma)}_{ml}(t)|^{2}\Big\rangle_{(\gamma)}}{\Big\langle \left(R_{j}^{(\alpha)}(t)-R^{(\beta)}_{j}(t)\right)\left(f^{(\beta)}_{j,m}(t) - f^{(\beta)}_{j,l}(t)\right)|\rho^{(\beta)}_{ml}(t)|^{2}\Big\rangle}_{(\beta)},    
\end{split}
\end{equation}
such that the quantum momentum is given by,
\begin{equation}\label{regularisation_2}
    \mathcal{P}^{(\alpha)}_{j,ml}(t) = \frac{\hbar}{2\sigma_{j}(t)}\frac{1}{x_{j,ml}^{(\alpha)}(t)}.    
\end{equation}
The benefit of this definition is that we may now identify the divergence as a $\frac{1}{x}$ divergence at $x=0$ and presuming that $x_{j,ml}^{(\alpha)}(t)$ is a smooth function of the BO momentum, nuclear positions and coherence, this divergence should be approached smoothly for an infinitesimal time step. It also provides the benefit that any regularised function replacing $\frac{1}{x}$ will contain a dimensionless regularisation parameter. For our later analysis we choose to regularise the quantum momentum via,
\begin{equation}\label{regularisation_3}
    \mathcal{P}^{(\alpha)}_{j,ml}(t) = \frac{\hbar}{2\sigma_{j}(t)}\frac{x_{j,ml}^{(\alpha)}(t)}{\left(x_{j,ml}^{(\alpha)}(t)^{2}+\epsilon^{2}\right)},
\end{equation}
where $\epsilon$ is a parameter that is chosen much smaller than the typical values of $x_{j,ml}^{(\alpha)}(t)$. 
Now the quantum momentum smoothly interpolates between its maximum value defined as $\pm\frac{\hbar}{2\epsilon}\frac{1}{2\sigma_{j}(t)}$ and zero. 
A similar regularisation procedure may be used when regularising the division by velocity in the CTMQC-E algorithm, but this was not implemented. 
We will refer to this algorithm as Regularised CTMQC or in short, CTMQC-R. Once again it may be trivially coupled to the 
CTMQC-E algorithm, in which case we refer to this method as CTMQC-ER.  

\section{Computational Details}
We tested our implementation using the well-known Tully models I-IV\cite{tully_model}, which are a set of one dimensional models with a single nuclear degree of freedom and two adiabatic states. Given the low dimensionality of these models, exact results can be obtained using standard direct propagation techniques of the Schr\"{o}dinger equation. Initial conditions for our swarm of classical trajectories were obtained from a Wigner distribution 
corresponding to a nuclear wavepacket given by:
\begin{equation}\label{init_nuclear_wavefunc}
    \chi(R,0) = \frac{1}{(\pi\Sigma^{2})^{\frac{1}{4}}}\exp{\left[-\frac{(R-R_{0})^{2}}{2\Sigma^{2}}+ik_{0}(R-R_{0})\right]},
\end{equation} 
where in all cases we chose $\Sigma=\sqrt{2}$ [a.u]. The width parameter modulating the strength of the quantum momentum in Eq.~\eqref{traditional_quantum_momentum} was frozen (i.e., treated 
time-independent),  $\sigma_{j}(t)\!=\!\sigma$, and chosen such that the reconstructed initial nuclear density given by the sum of Gaussians centred on the initial sampled positions of the nuclei
reproduced the exact initial nuclear density in Eq.~\eqref{init_nuclear_wavefunc} as close as possible. All simulations were carried out for a swarm of 200 trajectories and a width of $\sigma = \sqrt{\frac{1}{5}}$. 
We found that populations and coherences were well-converged with just $200$ trajectories (see Fig. S1). We note that other choices of $\sigma =\frac{3}{10}$ [a.u] or $\sigma =\frac{1}{2}$ [a.u] yield similar results. 


Electronic populations were initialized in the ground state: $|C^{(\alpha)}_{0}(0)|^{2}=1\hspace{0.1cm}\forall\,\,\alpha$.
Nuclear dynamics were propagated using the Velocity-Verlet algorithm and the wavefunction coefficients were evolved using the RK4 algorithm. 
In all cases electronic and nuclear dynamics were evolved using the same time steps. This time step was $\Delta t = 10$ [as], 
except when testing convergence with time step we used $\Delta t\in \left(2, 10, 20, 30, 40, 50\right)$ [as]. 

Simulations were carried out for all three divergence procedures defined above. 
(i) CTMQC-DI and  CTMQC-EDI  using Eq.~\eqref{DI} for the intercepts; when the denominator in this expression went to zero we assigned a value of $0$ to that intercept. 
As argued above, this choice is arbitrary and has no effect on dynamics.  (ii) CTMQC-R and  CTMQC-ER  using Eq.~\eqref{regularisation_3} with $\epsilon=0.05$ so that 
the peak value of the quantum momentum matched that of the Cut-Off method. (iii) simulations with the standard Cut-Off method, here simply referred to as CTMQC and CTMQC-E, 
where we chose a cutoff for the denominator of the intercept term in Eq.~\eqref{old_y-intercept} and for the number of standard deviations from the centre of the quantum momentum 
such that they matched the current default values used in the g-CTMQC code: a cut-off radius of $10\sigma$, corresponding to a quantum momentum cut-off of $\frac{5\hbar}{\sigma}$\cite{g_CTMQC}. The denominator cut-off was $10^{-8}$ [a.u].

For Tully model IV we found that $\epsilon =0.05$ and a cut-off of $10\sigma$ resulted in the quantum momentum switching definitions too frequently, so we used $\epsilon =0.005$ 
and a cut-off corresponding to $1000\sigma$ instead. Model IV's values for the quantum momenta were generally higher than for the other models, so a ``divergence" of 
the quantum momentum was identified even when it was not the case.  

\section{Results}

In the following we present results for Tully model I, the Single Avoided Crossing, in the high momentum case ($k_{0}=25$ [a.u]).  
The results and conclusions drawn for this model were representative of the ones for all other Tully models, which are presented 
in the Supplementary Information (Figures S2-5). Before we compare the different divergence schemes we explain the dynamics observed 
in terms of the Ehrenfest and XF contributions to CTMQC.    




\begin{figure}[hbt!]
    \includegraphics[width =8cm, height = 12.8cm]{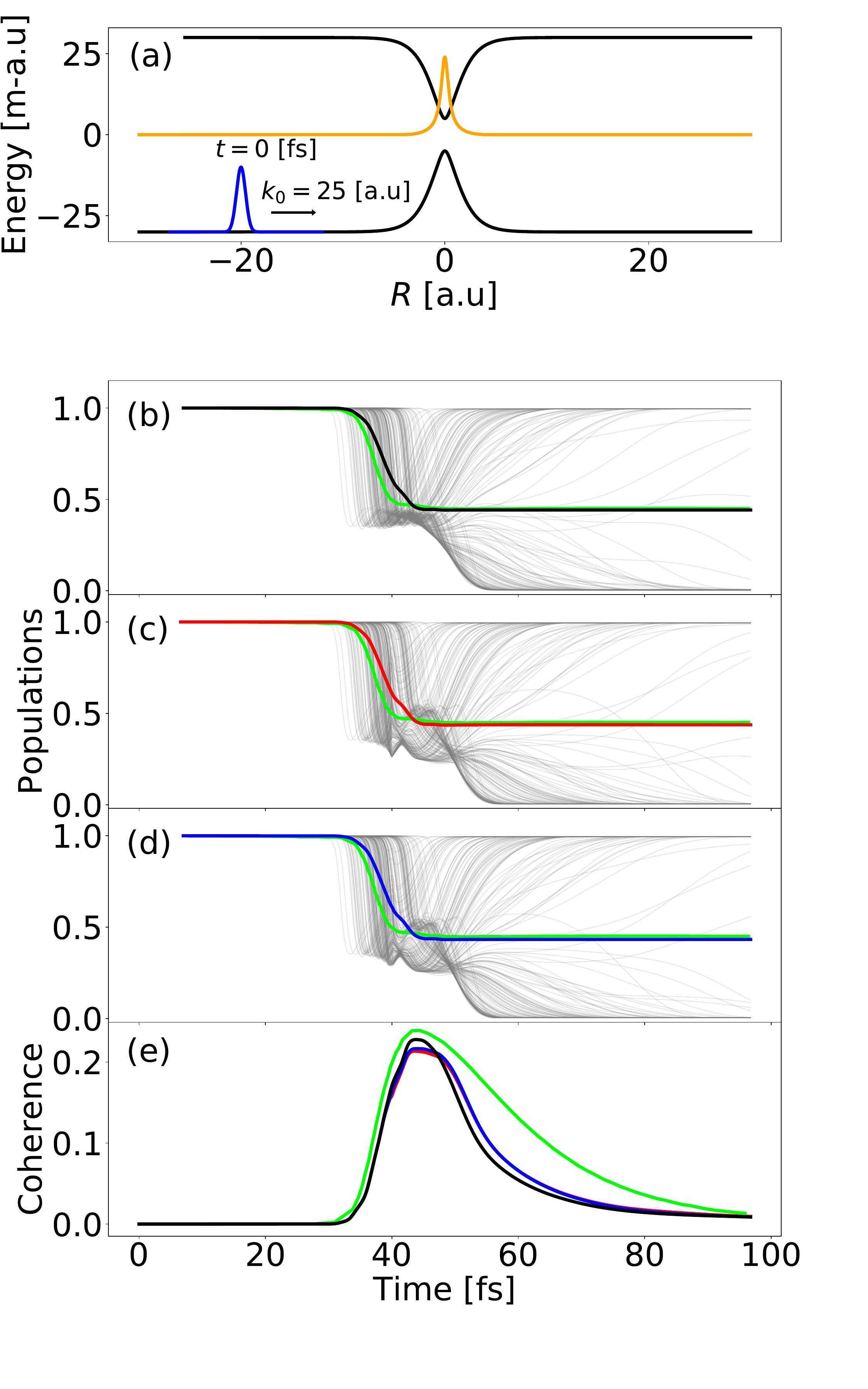}
    \vspace{-1.1cm}
    \
    \caption{Performance of different coupled trajectory mixed quantum classical (CTMQC) simulation methods on Tully model I. 
    (a) Sketch of the Single Avoided Crossing (SAC) potential energy surfaces (black), with the initial nuclear density (blue). The $d_{10}(R)$ NACV (orange) 
    is scaled by a factor of $\frac{1}{50}$. (b)-(d) Mean ground state populations, $\Big\langle|C_{0}^{(\alpha)}(t)|^{2}\Big\rangle_{(\alpha)}$,  in bold for the exact (green)\cite{gossel_benchmark} 
    and respective quantum momentum methods (b) CTMQC-EDI (black), (c) CTMQC-ER (red) and (d) CTMQC-E with Cut-Off method (blue), each plotted alongside 
    individual but coupled trajectories (grey). (e) Coherence, $\Big\langle|C_{0}^{(\alpha)}(t)|^{2}|C_{1}^{(\alpha)}(t)|^{2}\Big\rangle_{(\alpha)}$, using the same colour-coding as above. 
    Note the absence of spurious ``wiggles" in the populations of single trajectories at around 40 fs (avoided crossing region) for the CTMQC-EDI method in panel (b).}
    \label{fig:coh_and_pops}
\end{figure}

We may think of the dynamics in Tully I as going through three key regions defined by the potential energy surfaces and NACV in Figure \ref{fig:coh_and_pops}(a). In the region $R\in\left[-20, -5\right]$ [a.u] the coupling between states is weak, as quantified by the NACV, and this means that there is no population transfer on any of the trajectories because the dominating term in the evolution of the coefficients is the phase term in Eq.~\eqref{Ehrenfest_contribution}. This explains why the black line in Figure \ref{fig:coh_and_pops}(b) is initially flat for the first 30 [fs]. Since the system is initialised with the full population in the ground state, the XF contributions of Eq.~\eqref{XF_contribution} also vanish, since there is no electronic coherence between states as can be seen initially in Figure \ref{fig:coh_and_pops}(e) on the black line. An analogous argument for the forces leads to the conclusion that the force acting on the nucleus derives entirely from the ground state in this region; $F^{(\alpha)}(t) \approx -\frac{\partial}{\partial R}E_{0}(R^{(\alpha)}(t))\hspace{0.1cm}\forall\,\, \alpha$, as can be derived from Eqs.~\eqref{Ehrenfest_contribution} and \eqref{XF_contribution}. 

In the region $R\in[-5,5]$ [a.u] the NACV is at its strongest, owing to the proximity of the potential energy surfaces, and thus we have a population transfer to the excited state for some trajectories mediated by the NACV contribution; the second term on the RHS of Eq.~\eqref{Ehrenfest_contribution}. We see this population transfer regime in Figure \ref{fig:coh_and_pops}(b) for the black line between $35$ and $45$ [fs], and notably this average population is representative of the individual contributions from the grey lines. In this region, the XF contributions are still small because the BO momentum contributions, $(f_{1}^{(\alpha)}(t)-f_{0}^{(\alpha)}(t))$, in Eq.~\eqref{XF_contribution} have not had time to accumulate. For this reason, we can think of this region as the one in which the building of coherence between states via the NACV is the dominant mechanism mediating population transfer. A corresponding increase in coherence for the black line in Figure \ref{fig:coh_and_pops}(e) is indicative of this behaviour. By a similar argument, the forces in this region are dominated by the mean contribution from each adiabatic state. 

In the region $R\in[5,\infty)$, the XF contributions come into play because the coherence between electronic states, $|\rho_{10}^{(\alpha)}(t)|^{2}$, is generally high and the BO momentum has had time to accumulate over the trajectory and now dominates over the diminished NACV contributions.  We see from Eq.~\eqref{XF_contribution} that this mechanism leads to decoherence at a rate proportional to the coherence between the electronic states. We see this desired decay of coherence in Figure \ref{fig:coh_and_pops}(e) after approximately 45 [fs] has elapsed. The role of the quantum momentum, $\mathcal{P}_{10}^{(\alpha)}(t)$, for this particular model can be understood from the fact that each contribution inherits the same sign for the BO momentum difference and that, given the flat potential energy surfaces in this region, it does not change sign. Therefore, the quantum momentum is involved in a ``self-organisation" process, in which trajectories will decohere to the excited state or the ground state depending on whether they trail or lead the intercept term of Eq.~\eqref{DI}, and depending on whether they exist on the positive or negative branch. It is important that a correct proportion of the trajectories will decohere to one state or the other; this is ensured by the spurious population transfer condition, Eq.~\eqref{zero_pop_transfer}, resulting in nearly unchanged populations in this region (black line in Figure \ref{fig:coh_and_pops}(b)). During this organising process,  the force contribution of Eq.~\eqref{XF_force} serves to push the nuclei towards the centre of the nuclear density, which eventually leads to a splitting of the trajectories of the swarm, as they follow forces deriving from the different potential energy surfaces once the electronic coherence has fully decayed, and the adiabatic approximation has been restored. Many of the same conclusions we draw here can be used to understand the general mechanisms that CTMQC describes, although we note that for example in this model we had flat potential 
energy surfaces in the region dominated by the quantum momentum, allowing us to attribute the sign of the decoherence term entirely to the sign of the quantum momentum; this would not be the case if the potential energy surfaces were well-separated but nonetheless had large gradients, since this would allow the BO momentum difference to change sign in this region.

All divergence methods reproduce the mean populations, Figure~\ref{fig:coh_and_pops}(b-d), and coherence, Figure~\ref{fig:coh_and_pops}(e), well compared with exact results. Additionally, the results are nearly identical regardless of the application of the energy correction. This property was noted of the Extended Coupling Region Tully model\cite{ctmqc-e}. In the Supplementary Information, Figures S2-5, we demonstrate that this is true of all of the Tully models for the initial nuclear momenta tested. However, the quantum momentum divergence, characterised by the denominator of Eq.~\eqref{old_y-intercept} going to zero, occurs at approximately $38$ [fs] in the Cut-Off and Regularisation methods, and this where the coherences in Figure \ref{fig:coh_and_pops}(e) from the different methods diverge from each other. For regularisation parameters that correspond to a higher cut-off to the quantum momentum, it is likely that one will observe a spurious dip in the coherences in this region; such behaviour was observed in Ref.\cite{gossel_benchmark}. We see a signature of this behaviour in the individual trajectories for our choice of regularisation parameters, 
as many of the trajectories jump at this divergence (thin grey lines in Figure \ref{fig:coh_and_pops}(c)-(d)). This undesirable behaviour is less pronounced if using the Cut-Off method compared with the Regularisation 
method but is still present. CTMQC-EDI (Figure \ref{fig:coh_and_pops}(b)) has the smoothest individual trajectories and has no signatures of the divergence apparent in the populations.

CTMQC-EDI has a very similar qualitative behaviour to the other two divergence methods despite its contrasting definition. We attribute this to the fact that CTMQC-EDI is only activated in regions where the BO momentum difference, $(\tilde{f}_{1}^{(\alpha)}(t)-\tilde{f}_{0}^{(\alpha)}(t))$, between adiabatic states contains trajectories of each sign. Generally in the Tully models, most trajectories eventually fall into the same branch after leaving the NAC region, in which case it behaves identically to the original definition of the quantum momentum containing the intercept term from Eq.~\eqref{old_y-intercept}. In fact for our presented model, all trajectories eventually fell into the positive branch for the $1,0$ index ordering. One would expect the behaviour of CTMQC-EDI compared with Regularisation and Cut-Off methods, to vary at long times if a reasonable proportion of trajectories occupied each branch, but it is not necessarily true that one method would perform better than another in this circumstance. However, given that divergences of the quantum momentum may only occur if trajectories exist in both branches, the Regularisation and Cut-Off methods would be more subject to numerical instabilities.

\begin{figure}[htb!]
    \centering
      \includegraphics[width = 0.45\textwidth]{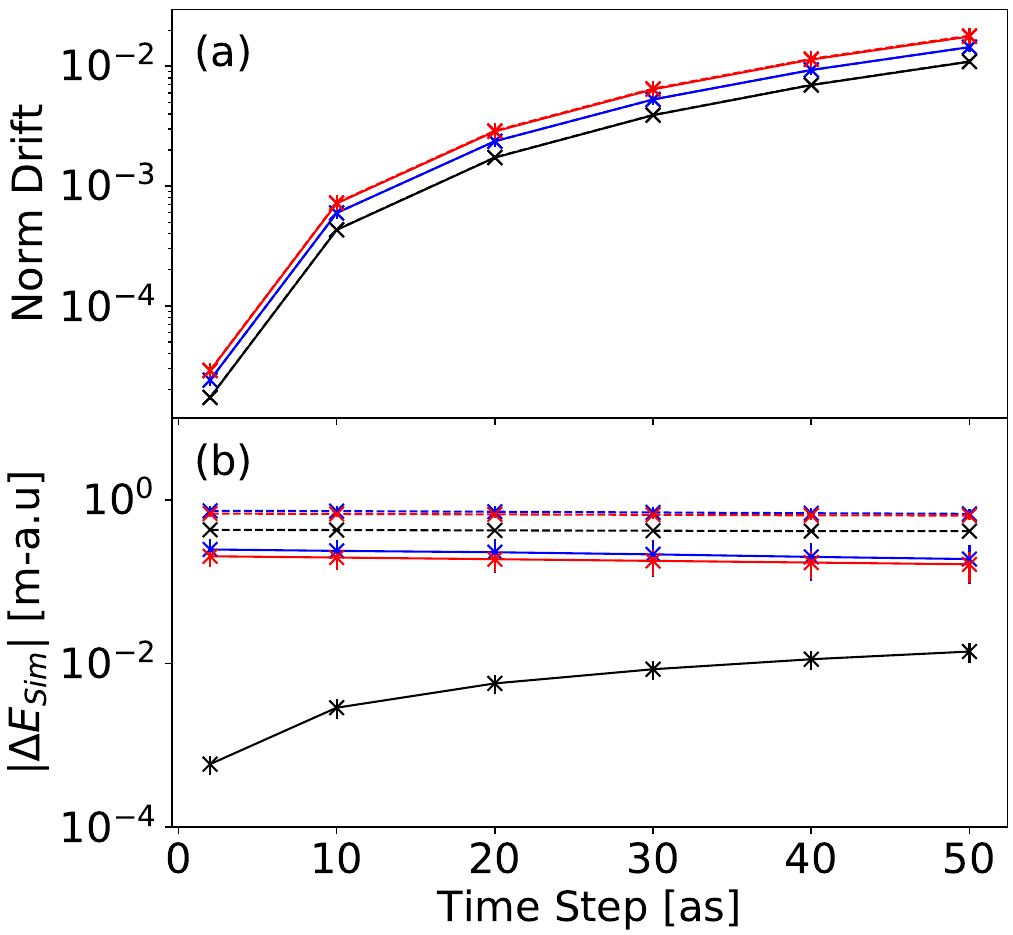}
    \caption{Norm and total energy conservation of different CTMQC methods against integration time step, for Tully model I. Deviations of electronic wavefunction norm from unity (a) 
    and total energy conservation (b), $\Delta E_{sim} \equiv\Big\langle E^{(\alpha)}(t_{final}) - E^{(\alpha)}(0)\Big\rangle_{(\alpha)}$, where $E^\alpha$ is given by Eq.~\eqref{energy_expression}. 
    Standard deviations were computed from 10 initial seeds for the Wigner distribution corresponding to Eq.~\eqref{init_nuclear_wavefunc} and are indicated by error bars but are small and may not
    be visible. Colour coding is as follows: CTMQC-EDI (black solid line), CTMQC-ER (red solid line), CTMQC-E with Cut-Off method (blue sold line), CTMQC-DI (black dashed line), CTMQC-R (red dashed line) and CTMQC with Cut-Off method
    (blue dashed line). When the energy correction is applied the electronic wavefunction is renormalised to decouple norm and energy drift.}
    \label{fig:norm_ener_plot}
\end{figure}

Figure \ref{fig:norm_ener_plot}(a) demonstrates that all algorithms perform similarly in terms of norm conservation, which improves with time step. The results are unaffected by the use of the energy correction algorithm. In Figure \ref{fig:norm_ener_plot}(b) absolute energy drift is improved by the CTMQC-E algorithm for all methods, while leaving the norm conservation unchanged.  Yet the improvement in energy conservation obtained with CTMQC-EDI is two orders of magnitude better than with the other intercept methods. Importantly, the energy conservation can be systematically improved by decreasing the integration time step, whereas no such improvement is obtained with the other two intercept methods. Moreover, in CTMQC-EDI, the performance in energy (and norm) conservation is least sensitive to the initial conditions chosen (smallest error bars in Figure \ref{fig:norm_ener_plot}), suggesting of the high numerical stability and robustness of this method.

\begin{figure}
    \centering
    \includegraphics[width = 0.45\textwidth]{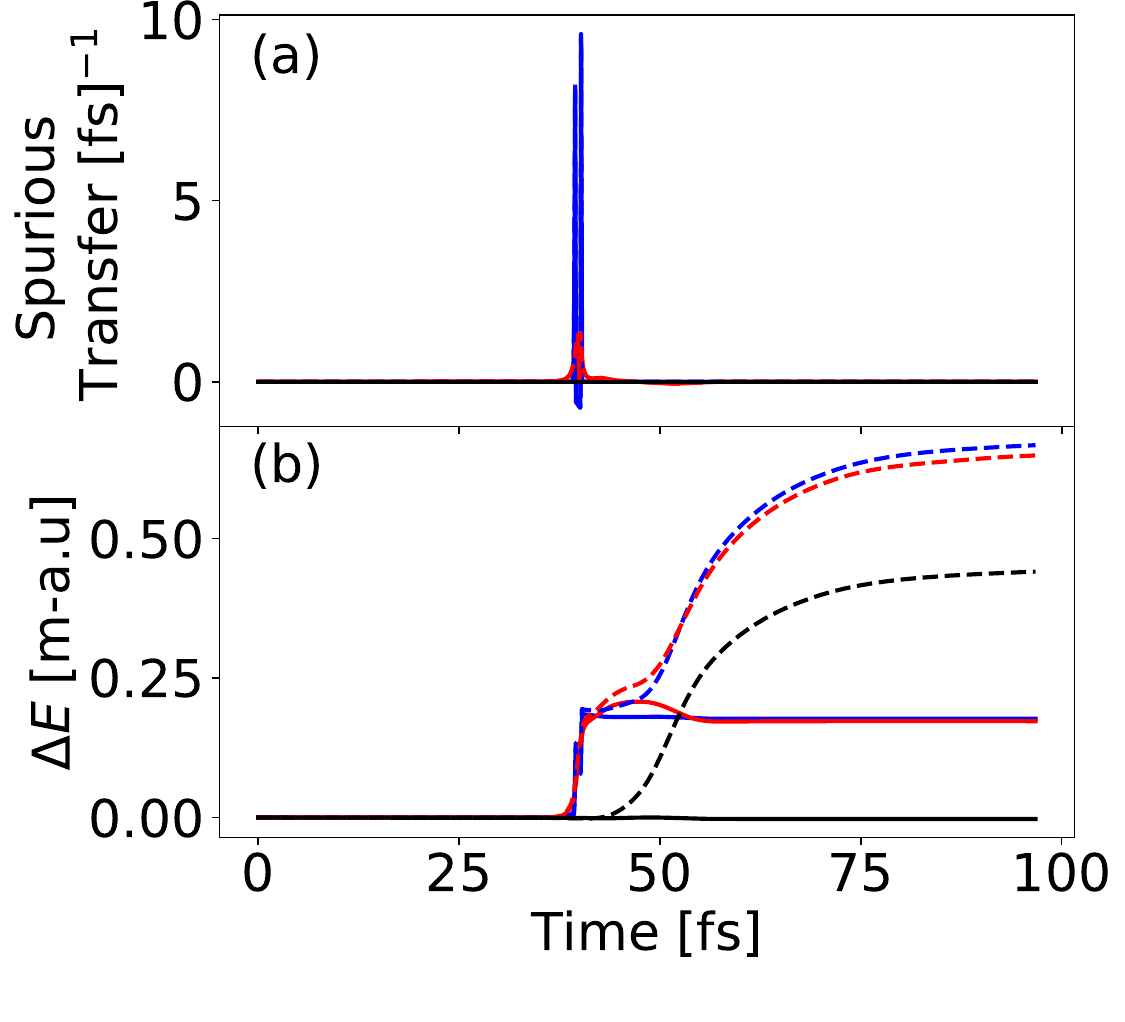}
    \caption{
    Spurious population transfer and total energy conservation of different CTMQC methods against time, for Tully model I.
    (a) Spurious population transfer indicator defined as $N_{traj}\Big\langle\frac{2}{M}\mathcal{P}_{10}^{(\alpha)}(t)(f_{1}^{(\alpha)}(t)-f_{0}^{(\alpha)}(t))|\rho_{10}^{(\alpha)(t)}|^{2}\Big\rangle_{(\alpha)}$
    and (b) trajectory-averaged energy drift defined as $\Delta E(t)\equiv\Big\langle E^{(\alpha)}(t)-E^{(\alpha)}(0)\Big\rangle_{(\alpha)}$ with $E^\alpha$ given by Eq.~\eqref{energy_expression}. 
     Colour code and line styles as in Figure~2.} 
    \label{fig:ener_spurious_trans}
\end{figure}


Figure \ref{fig:ener_spurious_trans}(a) demonstrates the reason for the improved energy conservation using CTMQC-EDI observed in Figure \ref{fig:norm_ener_plot}(b). At $38$ [fs], the intercept of the quantum momentum, Eq.~\eqref{old_y-intercept}, diverges. In the Cut-Off method the definition of the intercept is switched to Eq.~\eqref{gaussian_intercept} and in the Regularisation method, Eq.~\eqref{regularisation_3}, the regularisation becomes active. Both procedures lead to a violation of Eq.~\eqref{zero_pop_transfer} or, equivalently, of Eq.~\eqref{original_pop_transfer} and spurious population transfer at  $38$ [fs] (corresponding to peaks in Figure \ref{fig:ener_spurious_trans}(a)). By contrast, in CTMQC-EDI, the intercepts defined by Eq.~\eqref{DI} remain finite and Eq.\eqref{zero_pop_transfer} remains fulfilled at any time because of the stricter conditions defined 
by Eqs.~\eqref{zero_pop_transfer_plus}-\eqref{zero_pop_transfer_minus}.
Notice that the spurious population condition, on the left hand side of Eq.~\eqref{original_pop_transfer}, appears in the expression for change in total energy, given by the first term on the right hand side of 
Eq.~\eqref{Energy_deviation_XF}, and that conservation of total energy in the CTMQC-E methods is guaranteed only  if these terms are zero. Since they are no longer zero when switching definitions via the Cut-Off method or by interpolating using the Regularisation method, total energy is no longer conserved (see Figure \ref{fig:ener_spurious_trans} (b) solid blue and red lines respectively at $38$ [fs]). Hence, The -E extension
to the algorithms cures the energy conservation only prior and after this point in time where no definition switching or regularisation occurs and Eq.~\eqref{Energy_deviation_XF} is fulfilled (Figure \ref{fig:ener_spurious_trans}(b), compare solid with dashed lines).  By contrast, CTMQC-EDI does not suffer from this problem and the total energy remains conserved at any time (Figure 3(b), solid black line).  This also explains why we do not see an improvement in the energy conservation with time step (Figure \ref{fig:norm_ener_plot}(b)) despite the energy correction for Regularisation and Cut-Off methods; in these cases the vast majority of energy drift occurs due to the spurious population transfer, which occurs regardless of the time step used as the dynamics are well converged even for 50 [as] time steps.
%
%
The exact extent to which spurious population transfer occurs depends on a variety of factors, but principally on what defines the cut-off to the quantum momentum. If the cut-off is increased, then the spurious nature of the divergence can be greatly reduced, and the corresponding jump that would be observed in the energy would also be reduced. This is however merely due to the fact that one does not switch (or interpolate to in the case of the Regularisation method) to a non-conserving definition of the quantum momentum as often, and therefore runs the risk of causing increasingly unstable dynamics as the quantum momentum is allowed to increase to the new threshold, which is in principle arbitrarily large compared to the other terms contributing to the population transfer. There is always therefore an internal tension between conserving energy and stabilising the dynamics. CTMQC-EDI on the other hand always satisfies the 
spurious population transfer condition Eq.~\eqref{zero_pop_transfer} with deviations never exceeding $10^{-12}$ [fs]$^{-1}$.
 \section{Conclusion}
In summary, we proposed a new treatment of the quantum momentum in the CTMQC algorithm via a redefinition which removed the need for switching definitions or regularisation of the quantum momentum. We have proven the new method, CTMQC-DI, has the same desirable qualities as the contemporary methods, namely norm conservation and analytic continuity of forces and populations, whilst eliminating spurious population transfer. CTMQC-DI also shows good agreement with exact quantum mechanics for populations and coherences. Just like CTMQC, the algorithm is deterministic; nuclear dynamics are continuous and follow forces derived directly from the electronic wavefunction instead of a surface wavefunction, while electronic dynamics are also continuous at all times. Once coupled to the recently developed CTMQC-E algorithm, it was shown that the energy conservation was significantly improved over previous methods and was the only method that reliably improved with decreasing time step. Energy non-conservation is still a possibility if the kinetic energy of the system goes to zero on a given trajectory (see the right hand side of Eq.~\eqref{redef_BO_momentum}), but this becomes more infrequent the larger the system. For this reason, we expect that the advantage with respect to energy conservation incurred by the redefinition will improve the larger the system size. CTMQC-EDI therefore shows great promise as a non-adiabatic dynamics technique for molecular \cite{C6FD00107F} and large scale quantum systems. 
Future efforts will be focused on implementing CTMQC-EDI for simulating non-adiabatic processes in nanoscale molecular materials 
and comparing the results with existing non-adiabatic dynamics methods such as 
fragment orbital-based surface hopping (FOB-SH)\cite{C9CP04770K, carof2017detailed, giannini2019quantum, doi:10.1021/acs.jpclett.2c01928}.

\section{Acknowledgements}
A.D. was supported by a EPSRC DTP PhD studentship.

\bibliography{Refs_new_2}

\bibliographystyle{apsrev}

\end{document}